\documentclass{article}

\newcommand{\keywords}[1]{\vspace{1em}\noindent\textbf{Keywords:} #1}

\usepackage{amsmath,amssymb,amsfonts}%
\usepackage[title]{appendix}%

\begin{document}

\title{Proof of Heisenberg's Error-Disturbance Relation for Individual Measurements}

\author{Seiji Kosugi\footnote{Independent Researcher, Yokohama, Japan}\footnote{kosugi0507@gmail.com}}

\maketitle

\begin{abstract}
Heisenberg originally envisioned the error-disturbance relation under the premise that the state after measurement must remain consistent with the Kennard-Robertson uncertainty relation. 
This implies that the measurement error must be formulated via a posterior observable $\hat{x}_t$, rather than a prior observable $\hat{x}_0$, because the posterior measurement error determines the post-measurement uncertainty of the electron. Building upon the preliminary conceptual foundation reported in \cite{SK1}, this paper presents a rigorous derivation of the error-disturbance principle formulated from the unitary transformation equations of observables. 
Each readout $X$ of a posterior probe observable $\hat{X}_t$ completes a single measurement event, producing a specific conditional object state.
The Kennard-Robertson uncertainty relation must hold for these states. 
Specifically, we show that the uncertainty relation between the error $\epsilon_{X}(\hat{x}_t)$ and the disturbance $\eta_{X}(\hat{p}_0)$ holds strictly at the level of individual measurements, characterized by the specific readout $X$.
We also verify that conventional error-disturbance relations, where the errors are evaluated by averaging over the unconditioned state $|\Psi_t \rangle$, hold true. Our results provide a refined theoretical basis for understanding the fundamental trade-off in individual measurement outcomes.
\end{abstract}


\section{Introduction}\label{sec1}

In 1927, Heisenberg claimed that the error-disturbance relation (EDR) $\epsilon\eta \sim h$ holds between the position measurement error $\epsilon$ of an electron and the momentum disturbance $\eta$ on the basis of his famous $\gamma$-ray microscope thought experiment\cite{WHZP}. 
In that paper, he states the following; Suppose that the momentum of a free electron is precisely known, while its position is completely unknown.
A subsequent measurement of the position is performed with a very small error $\epsilon$.
Then, if the electron momentum were not altered drastically by the position measurement, the Kennard--Robertson uncertainty relation \cite{Kennard, Robertson} $\sigma(\hat{x}_t)\sigma(\hat{p}_t) \geq \hbar /2$ would fail to hold for the post-measurement state, where $\sigma(\hat{A})$ denotes the standard deviation of an observable $\hat{A}$.
Thus, quantum mechanics would be impossible.
From Heisenberg' reasoning above mentioned, clearly, he considered the error $\epsilon$ determines the uncertainty of the electron position immediately after the measurement.
This interpretation is also supported by the fact that Heisenberg concluded that the uncertainty relation to be valid for the electron motion \textit{after} the experiment in his $\gamma$-ray microscope thought experiment \cite{WHPP}.
As previously discussed in our initial arXiv preprint \cite{SK1}, Heisenberg's original motivation for introducing the error-disturbance relation rested on the requirement that the particle after measurement must still satisfy the Kennard-Robertson uncertainty relation.

Ozawa pointed out that there exists a position measurement where the error $\epsilon$ is zero while $\eta < \infty$ \cite{MOPL1}.
Therefore, the relation $\epsilon\eta \sim h$ would be invalid for such a measurement. Subsequently, many experimental tests of Heisenberg's EDR have been performed \cite{Nat, ScR, FK1, LAR, MMW, MR}.
These experimental tests demonstrated that the relation
\begin{equation}
  \epsilon(\hat{x}_0) \eta(\hat{y}_0) \geq \frac{1}{2}|\langle \phi_0 | [\hat{x}_0,\hat{y}_0] | \phi_0 \rangle |
\end{equation}
between the measurement error $\epsilon(\hat{x}_0)$ of a \textit{prior} observable $\hat{x}_0$ and the disturbance $\eta(\hat{y}_0)$ of another observable $\hat{y}_0$ caused by the $\hat{x}_0$ measurement can be violated, where $| \phi_0 \rangle$ is the state vector of the object immediately before the measurement.

Do these findings demonstrate that Heisenberg's original argument was incorrect? The key is that Heisenberg's original EDR is not the relation (1).
While inequality (1) does not always hold, in Heisenberg's original reasoning, the error of the position measurement is understood as determining the uncertainty of the electron position \textit{after} the measurement.
Therefore, the measurement error in Heisenberg's EDR must be that which determines the uncertainty in the electron position immediately after the measurement. As demonstrated below, it is the measurement error $\epsilon(\hat{x}_t)$ of the posterior observable $\hat{x}_t$, not the measurement error $\epsilon(\hat{x}_0)$ of the prior observable $\hat{x}_0$, that dictates the uncertainty in the electron position immediately after the measurement [see Eq. (\ref{ersd})].
Consequently, Heisenberg's original EDR must be formulated as a trade-off between the posterior error $\epsilon(\hat{x}_t)$ and the disturbance $\eta(\hat{y}_0)$:

\begin{equation}
  \epsilon(\hat{x}_t) \eta(\hat{y}_0) \geq \frac{1}{2}|\langle \phi_0|\langle \xi_0 | [\hat{x}_t,\hat{y}_t] | \phi_0 \rangle |\xi_0 \rangle|,
\end{equation}
where $| \xi_0 \rangle$ is the initial probe state vector, and we write the tensor product $|\phi_0\rangle \otimes |\xi_0 \rangle$ as $|\phi_0 \rangle |\xi_0 \rangle$.
Although Heisenberg demonstrated in a few thought experiments \cite{WHPP} that the uncertainty relation (2) was valid, it has never been rigorously proven to hold in general.
As will be shown, this relation holds universally.

Each readout $X$ of a posterior probe observable $\hat{X}_t$ completes a single measurement, producing a specific conditional object state.
Then, the uncertainty relation between $\epsilon_{X}(\hat{x}_t)$ and $\eta_{X}(\hat{p}_0)$ must hold for individual measurements characterized by the readout $X$. In the present paper, we significantly extend this point of view by taking the unitary transformation of observables as our fundamental starting point. We rigorously show that the uncertainty relation between $\epsilon_{X}(\hat{x}_t)$ and $\eta_{X}(\hat{p}_0)$ holds across all realized outcomes $X$.

\section{Measurement model}\label{sec2}

We consider the measurement of an observable $\hat{\sigma}_t$ of a microscopic object.
The object interacts with a probe (part of the apparatus) over the time interval $(0,t)$. Let $\hat{U}$ be a unitary operator representing the time evolution of the composite object-probe system during this interval.
The post-interaction object and probe observables, $\hat{\sigma}_t$ and $\hat{X}_t$, are given in the Heisenberg picture by $\hat{U}^{\dagger} (\hat{\sigma}_0 \otimes \hat{I}) \hat{U}$ and $\hat{U}^{\dagger} (\hat{I} \otimes \hat{X}_0) \hat{U}$, respectively. Hereafter, we abbreviate $\hat{\sigma}_0 \otimes \hat{I}$ as $\hat{\sigma}_0$.
After the interaction, the probe observable $\hat{X}_t$ is measured by a secondary apparatus.
Using the readout value $X$ thus obtained, the value of $\hat{\sigma}_t$ is estimated.
We assume that $\hat{X}_t$ can be measured precisely without further interaction between the secondary apparatus and the object system.
The probe is prepared in a fixed initial state $|\xi_0 \rangle$.
In Heisenberg's original case, $\hat{\sigma}_t$ and $\hat{X}_t$ represent the post-interaction position operators of the object and probe; here, however, we treat a more general setup.

Let $|\Phi_{\sigma, X}\rangle$ be a simultaneous eigenvector of $\hat{\sigma}_t$ and $\hat{X}_t$:
\begin{equation}
\hat{\sigma}_t |\Phi_{\sigma, X}\rangle = \sigma |\Phi_{\sigma, X}\rangle, \quad \hat{X}_t |\Phi_{\sigma, X}\rangle = X |\Phi_{\sigma, X}\rangle. \nonumber
\end{equation}
Since $\hat{\sigma}_t = \hat{U}^\dagger \hat{\sigma}_0 \hat{U}$ and $\hat{X}_t = \hat{U}^\dagger \hat{X}_0 \hat{U}$, it follows that
\begin{equation}
\hat{\sigma}_0 \hat{U} |\Phi_{\sigma, X}\rangle = \sigma \hat{U} |\Phi_{\sigma, X}\rangle, \quad \hat{X}_0 \hat{U} |\Phi_{\sigma, X}\rangle = X \hat{U} |\Phi_{\sigma, X}\rangle. \nonumber
\end{equation}
Therefore, $\hat{U} |\Phi_{\sigma, X}\rangle$ is a simultaneous eigenvector of $\hat{\sigma}_0$ and $\hat{X}_0$ with eigenvalues $\sigma$ and $X$:
\begin{equation}
\hat{U} |\Phi_{\sigma, X}\rangle = |\hat{\sigma}_0 = \sigma\rangle |\hat{X}_0 = X \rangle,
\end{equation}
where $|\hat{\sigma}_0 = \sigma\rangle$ ($|\hat{X}_0 = X \rangle$) is an eigenvector of $\hat{\sigma}_0$ ($\hat{X}_0$).
Equation (3) indicates that the state $|\Phi_{\sigma, X}\rangle$ transforms into $|\hat{\sigma}_0=\sigma\rangle |\hat{X}_0=X \rangle$ via the interaction.
Because a unitary operator defines a transformation between orthonormal bases in Hilbert space, the pre-measurement state $|\Phi_{\sigma, X}\rangle$ can be expressed as a product state $|\hat{x}_0=x_0 \rangle |\hat{\Sigma}_0=\Sigma_0 \rangle$ of the eigenvectors of an object observable $\hat{x}_0$ and a probe observable $\hat{\Sigma}_0$: 
\begin{align}
   \hat{U}|\hat{x}_0=x_0 \rangle |\hat{\Sigma}_0=\Sigma_0 \rangle = |\hat{\sigma}_0=\sigma \rangle |\hat{X}_0=X \rangle, \\
   \hat{x}_0 |\hat{x}_0=x_0 \rangle = x_0 |\hat{x}_0=x_0 \rangle,\quad \hat{\Sigma}_0 |\hat{\Sigma}_0=\Sigma_0 \rangle = \Sigma_0 |\hat{\Sigma}_0=\Sigma_0 \rangle,  \\
   \hat{\sigma}_0 |\hat{\sigma}_0=\sigma \rangle = \sigma |\hat{\sigma}_0=\sigma \rangle,\quad \hat{X}_0 |\hat{X}_0=X \rangle = X |\hat{X}_0=X \rangle.
\end{align}
Where unambiguous, we abbreviate $|\hat{x}_0=x_0\rangle$ as $|x_0 \rangle$.


\section{$\hat{U}|x_0 \rangle |\Sigma_0 \rangle = |\sigma \rangle |X \rangle$}\label{sec3}

We first consider continuous eigenvalues $x_0, \Sigma_0, \sigma$, and $X$.
As discussed in Sec. 1, Heisenberg sought to confirm that the Kennard--Robertson uncertainty relation holds for the posterior object state.
Because each measurement outcome determines a distinct state, the relation must hold across all realized outcomes.

Let $|\phi_0 \rangle |\xi_0 \rangle$ be the initial state of the composite system, and let $|\Psi_t \rangle = \hat{U}|\phi_0 \rangle |\xi_0 \rangle$ be the post-interaction state.
Using the wavefunction $\Psi_t(\sigma,X) = \langle \sigma|\langle X|\hat{U}|\phi_0 \rangle |\xi_0 \rangle$ and completeness $\int d\sigma dX |\sigma\rangle|X\rangle\langle\sigma|\langle X|$, we write
\begin{align}
   |\Psi_t \rangle \equiv \hat{U}|\phi_0 \rangle |\xi_0 \rangle = \int |\sigma\rangle|X \rangle \Psi_t(\sigma,X)d\sigma dX.
\end{align}
The overall measurement error for estimating $\hat{\sigma}_t$ using a measurement value $F(X)$ is
\begin{align}
   \epsilon^2(\hat{\sigma}_t) = \langle \phi_0| \langle \xi_0|\{F(\hat{X}_t)-\hat{\sigma}_t \}^2 |\phi_0\rangle |\xi_0 \rangle = \langle \Psi_t|\{F(\hat{X}_0)-\hat{\sigma}_0 \}^2 |\Psi_t \rangle.
\end{align}
Let $|\Psi_t \rangle_X$ be the conditional state immediately following a projective readout $X$ of $\hat{X}_t$:
\begin{align}
   |\Psi_t \rangle_{X} &= \int |\sigma\rangle|X \rangle \Psi_t(\sigma,X)d\sigma \equiv |\phi_t \rangle _{X} |X \rangle, \\
   |\Psi_t \rangle &= \int \sqrt{P(X)}|\Psi_t \rangle_{X}dX,
\end{align}
where $P(X)$ is the probability density for obtaining readout $X$.

The conditioned error $\epsilon_X(\hat{\sigma}_t)$ then satisfies
\begin{align}
\epsilon_{X}^2(\hat{\sigma}_t) = {}_X\langle \Psi_t|\{F(\hat{X}_0)-\hat{\sigma}_0 \}^2 | \Psi_t \rangle_X \nonumber = \int \{ F(X)-\sigma \}^2 |\langle \sigma|\phi_t \rangle _{X}|^2 d\sigma \, \langle X|X \rangle.
\end{align}
To resolve the formal divergence $\langle X|X \rangle = \delta(0)$, we employ the standard eigendifferential $|X \rangle_\epsilon$ \cite{MESSIAH}:
\begin{align}
   |\Psi_t \rangle_{X} = |\phi_t \rangle _{X}|X \rangle_{\epsilon}, \quad |X \rangle_\epsilon &\equiv \frac{1}{\sqrt{\epsilon}} \int_{X-\epsilon/2}^{X+\epsilon/2} |X^{\prime} \rangle dX^{\prime},
\end{align}
and take $\lim_{\epsilon \to 0}$ at the end of the evaluation.
For any $\epsilon > 0$, ${}_\epsilon \langle X |X \rangle_\epsilon = 1$ and $\lim_{\epsilon \to 0} {}_\epsilon \langle X |F(\hat{X}_0) | X \rangle_\epsilon = F(X)$. Here, the object state $\vert{}\phi_t\rangle_X$ is normalized (i.e., ${}_X\langle \phi_t \vert{} \phi_t \rangle_X = 1$), and the state of the total system also satisfies the normalization condition $\langle \Psi_t \vert{} \Psi_t \rangle = 1$. 

 Thus,
\begin{align}
   \epsilon_{X}^2(\hat{\sigma}_t) &= {}_X\langle \Psi_t|\{ \hat{\sigma}_0-F(X) \}^2 | \Psi_t \rangle_X \nonumber \\
 &= {}_X\langle \Psi_t|\{\hat{\sigma}_0-\langle \hat{\sigma}_0 \rangle_{X} \}^2 | \Psi_t \rangle_X + \{\langle \hat{\sigma}_0 \rangle_{X}-F(X) \}^2, \label{ersd}
\end{align}
where $\langle \hat{\sigma}_0 \rangle_{X} \equiv {}_X\langle \Psi_t|\hat{\sigma}_0 | \Psi_t \rangle_X$.
Equation (\ref{ersd}) shows that the measurement error is minimized when $F(X) = \langle \hat{\sigma}_0 \rangle_{X} \equiv \bar{F}(X)$, with the minimal value given by the standard deviation $\sigma_X(\hat{\sigma}_0)$:
\begin{equation}\label{sd2}
   \sigma_{X}^2(\hat{\sigma}_0) = {}_X\langle \Psi_t|\{ \hat{\sigma}_0-\langle \hat{\sigma}_0 \rangle_{X} \}^2| \Psi_t \rangle_X.
\end{equation}
This establishes $\epsilon_X(\hat{x}_t) \ge \sigma_X(\hat{x}_0)$, consistent with our previous work \cite{SK2}.

Next, let $\eta_X (\hat{y}_0)$ be the disturbance of $\hat{y}_0$ conditioned on readout $X$. Using $\hat{y}_t-\hat{y}_0 = \hat{U}^\dagger \{\hat{y}_0 - \hat{U}\hat{y}_0\hat{U}^\dagger \}\hat{U}$, we have
\begin{equation}
   \eta_{X}^2(\hat{y}_0) = {}_X\langle \Psi_t|\{\hat{y}_0-\hat{U}\hat{y}_0\hat{U}^\dagger \}^2 | \Psi_t \rangle_X,
\end{equation}
which yields the Robertson-type bound
\begin{equation}
   \sigma_X(\hat{\sigma}_0)\eta_X(\hat{y}_0) \ge \frac{1}{2}|{}_X\langle \Psi_t |[ \hat{\sigma}_0-\langle \hat{\sigma}_0 \rangle_{X},\hat{y}_0-\hat{U}\hat{y}_0\hat{U}^\dagger ] |\Psi_t \rangle_X|.
\end{equation}
Consider the cross-term
\begin{align}
&{}_{X}\langle \Psi_t | [\hat{\sigma}_0,\hat{U}\hat{y}_0 \hat{U}^{\dagger}]|\Psi_t \rangle_{X} =\int {}_{X}\langle \Psi_t |\sigma\rangle \langle \sigma | [\hat{\sigma}_0, \hat{U}\hat{y}_0 \hat{U}^{\dagger} ]|\sigma^{\prime}\rangle\langle \sigma^{\prime}|\Psi_t \rangle_X d\sigma d\sigma^{\prime}.
\end{align}
Defining
\begin{equation} 
I \equiv \langle \sigma|{}_\epsilon\langle X|\hat{\sigma}_0\hat{U}\hat{y}_0 \hat{U}^{\dagger}|\sigma^{\prime}\rangle |X \rangle_{\epsilon},
\end{equation}
and using $\hat{U}^{\dagger}|\sigma\rangle|X \rangle_{\epsilon} = |x_0\rangle|\Sigma_0\rangle_{\epsilon}$ and $\hat{U}^{\dagger}|\sigma^{\prime}\rangle|X \rangle_{\epsilon} = |x_0^{\prime}\rangle|\Sigma_0^{\prime}\rangle_{\epsilon}$, $I$ reduces to
\begin{equation} 
I = \sigma\langle x_0|\hat{y}_0|x_0^{\prime}\rangle ({}_\epsilon\langle \Sigma_0|\Sigma_0^{\prime}\rangle_{\epsilon}).
\end{equation}
Similarly, for
\begin{equation}
   J \equiv \langle \sigma|{}_\epsilon\langle X|\hat{U}\hat{y}_0 \hat{U}^{\dagger}\hat{\sigma}_0|\sigma^{\prime}\rangle |X \rangle_{\epsilon},
\end{equation}
we obtain
\begin{equation}
   J = \sigma^{\prime}\langle x_0|\hat{y}_0|x_0^{\prime}\rangle ({}_\epsilon\langle \Sigma_0|\Sigma_0^{\prime}\rangle_{\epsilon}).
\end{equation}
Because
\begin{align}
\lim_{\epsilon \to 0}{}_\epsilon \langle \Sigma_0 |\Sigma_0 ^{\prime}\rangle_\epsilon = \left\{
\begin{array}{ll}
1, & \Sigma_0 = \Sigma_0 ^{\prime}, \\
0, & \Sigma_0 \ne \Sigma_0 ^{\prime},
\end{array}
\right.
\end{align}
it follows that $I = J = 0$ whenever $\Sigma_0 \ne \Sigma_0^{\prime}$.

For $\Sigma_0 = \Sigma_0^{\prime}$, Eq. (4) describes an interaction mapping $(|x_0\rangle, |\Sigma_0\rangle) \to (|\sigma\rangle, |X\rangle)$.
In an indirect measurement, the inferable target value $\sigma$ must be uniquely determined by the initial probe parameter $\Sigma_0$ and final readout $X$.
If $(\Sigma_0, X) \mapsto \sigma$ were not single-valued, the measurement outcome would be ambiguous. Thus, $\Sigma_0 = \Sigma_0'$ forces $\sigma = \sigma'$ for a given $X$, guaranteeing $I = J$. \footnote{A rigorous mathematical proof showing that the map $(\Sigma_0, X) \mapsto \sigma$ is strictly single-valued under arbitrary unitary measurement dynamics will be presented in a forthcoming publication.}
Consequently,
\begin{equation}
   {}_X\langle \Psi_t | [\hat{\sigma}_0,\hat{U}\hat{y}_0 \hat{U}^{\dagger}]|\Psi_t \rangle_X = 0,
\end{equation}
which implies
\begin{equation}
   \sigma_X(\hat{\sigma}_0)\eta_X(\hat{y}_0) \ge \frac{1}{2}|{}_X\langle \Psi_t |[ \hat{\sigma}_0,\hat{y}_0 ] |\Psi_t \rangle_X|.
\end{equation}
Because $\epsilon_X(\hat{\sigma}_t) \ge \sigma_X(\hat{\sigma}_0)$ for any measurement value $F(X)$, we arrive at
\begin{equation}
\epsilon_X(\hat{\sigma}_t)\eta_X(\hat{y}_0) \ge \frac{1}{2}|{}_X\langle \Psi_t |[ \hat{\sigma}_0,\hat{y}_0 ] |\Psi_t \rangle_X|.
\end{equation}
Hence, Heisenberg's EDR holds for every outcome-conditioned measurement.

\section{$\hat{U}|x_i \rangle |\Sigma_0 \rangle = |\sigma_k \rangle |X \rangle$}\label{sec4}

We next examine the case where the object spectrum is discrete while the probe spectrum remains continuous:
\begin{align}
   \hat{U}|\hat{x}_0=x_i \rangle |\hat{\Sigma}_0=\Sigma_0 \rangle = |\hat{\sigma}_0=\sigma_k \rangle |\hat{X}_0=X \rangle, \\
   \hat{x}_0 |x_i \rangle = x_i |x_i \rangle, \quad \hat{\Sigma}_0 |\Sigma_0 \rangle = \Sigma_0 |\Sigma_0 \rangle, \quad \hat{\sigma}_0 |\sigma_k \rangle = \sigma_k |\sigma_k \rangle, \quad \hat{X}_0 |X \rangle = X |X \rangle. 
\end{align}

Because $\Sigma_0$ varies continuously, if $\sigma_k$ depended explicitly on $\Sigma_0$, its values would be continuous, contradicting discreteness. Thus, $\sigma_k$ depends solely on $x_i$:
\begin{equation}
\sigma_k = f(x_i).
\end{equation}

Similarly, $X$ must depend on both $x_i$ and $\Sigma_0$. If $X$ were independent of $x_i$, readout $X$ would contain no object information. Conversely, if $X$ depended only on $x_i$, then $X = g(\sigma_k)$, implying $\hat{X}_0 = g(\hat{\sigma}_0)$, which violates the assumption that $\hat{X}_0$ and $\hat{\sigma}_0$ act on distinct Hilbert spaces. Thus,
\begin{equation}
X = g(x_i,\Sigma_0).
\end{equation}

Following the derivation in Sec. 3, the optimal measurement value $F(X) = \langle \hat{\sigma}_0 \rangle_X$ minimizes $\epsilon_X(\hat{\sigma}_t)$ down to $\sigma_X(\hat{\sigma}_0)$, recovering Eq. (15).
Evaluating $I$ and $J$ as before:
\begin{align} 
I &\equiv \langle \sigma_k|{}_\epsilon\langle X|\hat{\sigma}_0\hat{U}\hat{y}_0 \hat{U}^{\dagger}|\sigma_{k^{\prime}}\rangle |X \rangle_{\epsilon} = \sigma_k\langle x_i|\hat{y}_0|x_{i^{\prime}}\rangle ({}_\epsilon\langle \Sigma_0|\Sigma_0^{\prime}\rangle_{\epsilon}), \\
J &\equiv \langle \sigma_k|{}_\epsilon\langle X|\hat{U}\hat{y}_0 \hat{U}^{\dagger}\hat{\sigma}_0|\sigma_{k^{\prime}}\rangle |X \rangle_{\epsilon} = \sigma_{k^{\prime}}\langle x_i|\hat{y}_0|x_{i^{\prime}}\rangle ({}_\epsilon\langle \Sigma_0|\Sigma_0^{\prime}\rangle_{\epsilon}).
\end{align}
Single-valuedness of $(\Sigma_0, X) \mapsto \sigma_k$ ensures that $\Sigma_0 = \Sigma_0'$ forces $\sigma_k = \sigma_{k^{\prime}}$, giving $I = J$.
Thus, Eq. (22) holds, confirming Heisenberg's EDR (24) for discrete object spectra.

\section{$\hat{U}|x_i \rangle |\Sigma_j \rangle = |\sigma_k \rangle |X_l \rangle$}\label{sec5}

Finally, consider fully discrete object and probe spectra:
\begin{align}
   \hat{U}|\hat{x}_0=x_i \rangle |\hat{\Sigma}_0=\Sigma_j \rangle = |\hat{\sigma}_0=\sigma_k \rangle |\hat{X}_0=X_l \rangle, \\
   \hat{x}_0 |x_i \rangle = x_i |x_i \rangle, \quad \hat{\Sigma}_0 |\Sigma_j \rangle = \Sigma_j |\Sigma_j \rangle, \quad  \hat{\sigma}_0 |\sigma_k \rangle = \sigma_k |\sigma_k \rangle, \quad \hat{X}_0 |X_l \rangle = X_l |X_l \rangle. 
\end{align}

Given a discrete outcome $X_l$ and measurement value $F(X_l)$, the conditional error is
\begin{equation}
\epsilon_{X_l}^2(\hat{\sigma}_t) = {}_{X_l}\langle \Psi_t| \{ F(\hat{X}_0)-\hat{\sigma}_0 \}^2 |\Psi_t \rangle_{X_l}.
\end{equation}
Setting $F(X_l) = {}_{X_l}\langle \Psi_t| \hat{\sigma}_0|\Psi_t \rangle_{X_l}$ minimizes the error to $\sigma_{X_l}(\hat{\sigma}_0)$, yielding
\begin{equation} 
   \sigma_{X_l}(\hat{\sigma}_0)\eta_{X_l}(\hat{y}_0) \ge \frac{1}{2}|{}_{X_l}\langle \Psi_t |[ \hat{\sigma}_0-\langle \hat{\sigma}_0 \rangle_{X_l},\hat{y}_0-\hat{U}\hat{y}_0\hat{U}^\dagger ] |\Psi_t \rangle_{X_l}|.
\end{equation}
The matrix elements evaluate to
\begin{align} 
   I &\equiv \langle \sigma_k|\langle X_l|\hat{\sigma}_0\hat{U}\hat{y}_0 \hat{U}^{\dagger}|\sigma_{k^{\prime}}\rangle |X_l \rangle = \sigma_k\langle x_i|\hat{y}_0|x_{i^{\prime}}\rangle \langle \Sigma_j|\Sigma_{j^{\prime}}\rangle, \\
   J &\equiv \langle \sigma_k|\langle X_l|\hat{U}\hat{y}_0 \hat{U}^{\dagger}\hat{\sigma}_0|\sigma_{k^{\prime}}\rangle |X_l \rangle = \sigma_{k^{\prime}}\langle x_i|\hat{y}_0|x_{i^{\prime}}\rangle \langle \Sigma_j|\Sigma_{j^{\prime}}\rangle.
\end{align}
If $j \ne j'$, $I = J = 0$. For $j = j'$, single-valuedness of $(\Sigma_j, X_l) \mapsto \sigma_k$ requires $\sigma_k = \sigma_{k^{\prime}}$, so $I = J$.
Thus, Eq. (22) holds, validating Heisenberg's EDR in discrete systems.

\section{Error-disturbance relations for the unconditioned state $\vert{}\Psi_t\rangle$}\label{sec6}

The EDRs established above apply to individual measurement with readout $X$.
However, in the conventional error-disturbance relations, the errors are evaluated by averaging over the unconditioned state $\vert\Psi_t\rangle$.
We now verify that the relation remains valid when errors are averaged over the unconditioned state $|\Psi_t\rangle$.

Let $\bar{\epsilon}_X(\hat{\sigma}_t)$ be the error associated with $\bar{F}(X) = {}_X\langle \Psi_t | \hat{\sigma}_0 | \Psi_t \rangle_X$. Then $\epsilon_X(\hat{\sigma}_t) \ge \bar{\epsilon}_X(\hat{\sigma}_t)$, and
\begin{align}
   \langle \Psi_t | \bar{F}(\hat{X}_0) | \Psi_t \rangle &= \int P(X)\bar{F}(X)dX \nonumber  = \int P(X) {}_X\langle \Psi_t | \hat{\sigma}_0 | \Psi_t \rangle_X dX  \nonumber \\ &= \langle \Psi_t | \hat{\sigma}_0 | \Psi_t \rangle. 
\end{align}
Thus, the measurement is globally unbiased:
\begin{equation}
\langle \phi_0 | \langle \xi_0 | \{ \bar{F}(\hat{X}_t) - \hat{\sigma}_t \} | \phi_0 \rangle | \xi_0 \rangle = 0 \quad \forall |\phi_0 \rangle.
\end{equation}
By Appleby's lemma \cite{App}, if an object-probe operator $\hat{A}$ satisfies $\langle \phi_0 |\langle \xi_0|\hat{A}|\phi_0 \rangle |\xi_0 \rangle = 0$ for all $|\phi_0 \rangle$, then
\begin{equation}
  \langle \phi_0|\langle \xi_0|\hat{A}|\phi_0^\prime \rangle |\xi_0 \rangle = 0 \quad \forall |\phi_0 \rangle, |\phi_0^\prime \rangle.
\end{equation}
Applying this to $\hat{A} = [\bar{F}(\hat{X}_t)-\hat{\sigma}_t, \hat{y}_0]$ gives
\begin{equation}
   \langle \phi_0| \langle \xi_0| [\bar{F}(\hat{X}_t)-\hat{\sigma}_t,\hat{y}_0]|\phi_0 \rangle |\xi_0 \rangle = 0.
\end{equation}
Therefore,
\begin{equation} 
 \epsilon(\hat{\sigma}_t)\eta(\hat{y}_0) \ge \bar{\epsilon}(\hat{\sigma}_t)\eta(\hat{y}_0) \ge \frac{1}{2}|\langle \phi_0| \langle \xi_0|[\bar{F}(\hat{X}_t)-\hat{\sigma}_t,\hat{y}_t-\hat{y}_0]|\phi_0 \rangle |\xi_0 \rangle|.
\end{equation}
Substituting Eq. (40) yields
\begin{equation} 
\epsilon(\hat{\sigma}_t)\eta(\hat{y}_0) \ge \frac{1}{2}|\langle \phi_0| \langle \xi_0|[\hat{\sigma}_t,\hat{y}_t]|\phi_0 \rangle |\xi_0 \rangle| = \frac{1}{2}|\langle \Psi_t|[\hat{\sigma}_0,\hat{y}_0]| \Psi_t \rangle|.
\end{equation}
This confirms that the unconditioned EDR holds universally.

\section{Discussion and Conclusion}\label{sec7}

Even if an object satisfies the Kennard--Robertson relation initially, it is not self-evident that it satisfies this relation after the measurement.
Heisenberg argued that the measurement error dictates the fluctuation in position of the electron after measurement.
Crucially, as shown here, this uncertainty corresponds to the error $\epsilon(\hat{x}_t)$ of the posterior observable $\hat{x}_t$, rather than $\epsilon(\hat{x}_0)$ of the prior observable $\hat{x}_0$.
Consequently, Heisenberg's EDR is represented by Eq. (2) rather than Eq. (1).
Although early quantum formulations treated $\hat{x}_0$ and $\hat{x}_t$ as identical via projection postulates \cite{VNMF}, settings where $\hat{x}_t \neq \hat{x}_0$ require a clear distinction.

Each readout $X$ completes a single measurement event, producing a specific conditional object state.
We have demonstrated that Heisenberg's EDR holds for these individual measurements. 

To illustrate the physical mechanism underlying Heisenberg's EDR, consider von Neumann's original interaction model \cite{VNMF}:
\begin{equation}
\hat{U} = \exp \left( -\frac{i}{\hbar} K \hat{x}_0 \hat{P}_0 t \right),
\end{equation}
where $\hat{P}_0$ is the probe momentum and $g_0 = Kt$. The interaction acts as
\begin{align}
\hat{U}|x_0 \rangle |X_0 \rangle &= |x \rangle |X \rangle, \quad x = x_0, \quad X = g_0 x_0 + X_0, \\
\hat{U}|p_0 \rangle |P_0 \rangle &= |p \rangle |P \rangle, \quad p = p_0 - g_0 P, \quad P = P_0 \quad (g_0 \neq 0).
\end{align}
Since $\hat{x}_0 = \frac{1}{g_0} \hat{X}_0 - \frac{1}{g_0} \hat{U} \hat{X}_0 \hat{U}^{\dagger}$, the minimal error is
\begin{equation}
   \epsilon_{X}^2(\hat{x}_t) = \frac{1}{g_0^2} \sigma_{X}^2(\hat{U} \hat{X}_0 \hat{U}^{\dagger}).
\end{equation}
Combined with $\hat{p}_t = \hat{p}_0 - g_0 \hat{P}_0$, which gives $\eta_{X}^2(\hat{p}_0) = g_0^2 ({}_X\langle \Psi_t | (\hat{U} \hat{P}_0 \hat{U}^\dagger)^2 | \Psi_t \rangle_X)$, we obtain
\begin{equation}
\epsilon_X(\hat{x}_t) \eta_X(\hat{p}_0) = \sigma_{X}(\hat{U} \hat{X}_0 \hat{U}^\dagger) \sigma_{X}(\hat{U} \hat{P}_0 \hat{U}^\dagger) = \frac{\hbar}{2}.
\end{equation}
Thus, the trade-off between error and disturbance directly reflects the standard-deviation trade-off between the conjugate probe observables $\hat{U}\hat{X}_0\hat{U}^\dagger$ and $\hat{U}\hat{P}_0\hat{U}^\dagger$.
This framework clarifies the conceptual foundations of Heisenberg's original trade-off relation.


\section*{Declarations}

Not applicable.


\begin{thebibliography}{99}

\bibitem{SK1} S. Kosugi, arXiv:1504.03779 [quant-ph].
\bibitem{WHZP} W. Heisenberg, Z. Phys. \textbf{43}, 172 (1927).
\bibitem{Kennard} E. H. Kennard, Z. Phys. \textbf{44}, 326 (1927).
\bibitem{Robertson} H. P. Robertson, Phys. Rev. \textbf{34}, 163 (1929).
\bibitem{WHPP} W. Heisenberg, \textit{The Physical Principles of the Quantum Theory} (University of Chicago Press, Chicago, 1930); reprinted by Dover, New York, 1967.
\bibitem{MOPL1} M. Ozawa, Phys. Lett. A \textbf{299}, 1 (2002).
\bibitem{Nat} J. Erhart, S. Sponar, G. Sulyok, G. Badurek, M. Ozawa, and Y. Hasegawa, Nat. Phys. \textbf{8}, 185 (2012).
\bibitem{ScR} S.-Y. Baek, F. Kaneda, M. Ozawa, and K. Edamatsu, Sci. Rep. \textbf{3}, 2221 (2013).
\bibitem{FK1} F. Kaneda, S.-Y. Baek, M. Ozawa, and K. Edamatsu, Phys. Rev. Lett. \textbf{112}, 020402 (2014).
\bibitem{LAR} L. A. Rozema, A. Darabi, D. H. Mahler, A. Hayat, Y. Soudagar, and A. M. Steinberg, Phys. Rev. Lett. \textbf{109}, 100404 (2012).
\bibitem{MMW} M. M. Weston, M. J. W. Hall, M. S. Palsson, H. M. Wiseman, and G. J. Pryde, Phys. Rev. Lett. \textbf{110}, 220402 (2013).
\bibitem{MR} M. Ringbauer, D. N. Biggerstaff, M. A. Broome, A. Fedrizzi, C. Branciard, and A. G. White, Phys. Rev. Lett. \textbf{112}, 020401 (2014).
\bibitem{MESSIAH} A. Messiah, \textit{Quantum Mechanics} (Dover Publications, New York, 1999).
\bibitem{SK2} S. Kosugi, Phys. Rev. A \textbf{82}, 022118 (2010).
\bibitem{App} D. M. Appleby, Int. J. Theor. Phys. \textbf{37}, 1491 (1998).
\bibitem{VNMF} J. von Neumann, \textit{Mathematical Foundations of Quantum Mechanics} (Princeton University Press, Princeton, NJ, 1955).
\end{thebibliography}
\end{document}